\documentclass[twocolumn,prd,amssymb,showpacs,nofootinbib,superscriptaddress,nofootinbib,amsmath]{revtex4-1}

\usepackage{graphicx}
\usepackage{dcolumn}
\usepackage{bm}
\usepackage{feynmf}
\usepackage{verbatim}
\usepackage{subfigure}

\newcommand{\p}{\partial×}
\newcommand{\tchi}{\tilde{\chi}×}
\newcommand{\ds}{\displaystyle×}
\newcommand{\tcs}{\langle \sigma v \rangle×}

\begin{document}

\title{Multi-State Dark Matter from Spherical Extra Dimensions}

\author{Peter T. Winslow}
\email{pwinslow@phas.ubc.ca}
\affiliation{Department of Physics and Astronomy, University of British Columbia, Vancouver, BC, V6T 1Z1, Canada}
\affiliation{Theory Group, TRIUMF, 4004 Wesbrook Mall, Vancouver, BC, V6T 2A3, Canada}
\author{Kris Sigurdson}
\email{krs@phas.ubc.ca}
\affiliation{Department of Physics and Astronomy, University of British Columbia, Vancouver, BC, V6T 1Z1, Canada}
\author{John N. Ng}
\email{misery@triumf.ca}
\affiliation{Theory Group, TRIUMF, 4004 Wesbrook Mall, Vancouver, BC, V6T 2A3, Canada}

\begin{abstract}
We demonstrate a new model which uses an ADD type braneworld scenario to produce a multi-state theory of dark matter. Compactification of the extra dimensions onto a sphere leads to the association of a single complex scalar in the bulk with multiple Kaluza-Klein towers in an effective four-dimensional theory. A mutually interacting multi-state theory of dark matter arises naturally within which the dark matter states are identified with the lightest Kaluza-Klein particles of fixed magnetic quantum number. These  states are protected from decay by a combination of a global $U(1)$ symmetry and the continuous rotational symmetry about the polar axis of the spherical geometry. We briefly discuss the relic abundance calculation and investigate the spin-independent elastic scattering off nucleons of the lightest and next-to-lightest dark matter states. \\ \\ \\
\end{abstract}

\maketitle

\section{Introduction}
The standard model (SM) of particle physics, while describing the results of collider experiments with unprecedented precision, lacks a suitable candidate for dark matter.  This is a problem that deserves urgent attention, as  cosmological observations have both measured the energy density of atoms and all other SM particles to be $\Omega_{\rm SM} h^2 \simeq 0.02$, and shown that a new type of \mbox{\emph{non-SM}} dark matter contributes $\Omega_{d} h^2 \simeq 0.11$ to the density of the Universe \cite{Dunkley:2008ie}. Corroborating evidence from astrophysics and structure formation suggests that dark matter, whatever it is, appears like non-relativistic cold dark matter (CDM) particles interacting at most weakly with SM particles (see, e.g., Refs.~\cite{Jungman:1995df,Bergstrom:2000pn,Bertone:2004pz,Hooper:2007qk} or, e.g., Ref.~\cite{D'Amico:2009df} for a pedagogical review). 

Most of the theoretical focus to date has been on minimal dark matter theories containing a single new stable elementary particle. However, while this is a reasonable conservative assumption, this need not be the case.  Furthermore, anomalies such as recent cosmic ray results \cite{:2008zzr,Adriani:2008zr,Abdo:2009zk} and long-observed annual modulation by the DAMA experiment \cite{Bernabei:2008yi} have sparked theoretical interest in a non-minimal dark sector.  It is worth investigating the possibility that the dark sector might not be so simple and may, in reality, be better described by a multi-particle set up (see, e.g., Refs.~\cite{Zurek:2008qg,Profumo:2009tb}).  

Extra dimensional models have been studied extensively as a possibility for new physics in both particle physics and cosmology. Interest in these models has been fueled by, amongst other things, their ability to accommodate Higgsless electroweak symmetry breaking \cite{Csaki:2003zu,Csaki:2003sh} and provide potential solutions to the hierarchy problem \cite{ArkaniHamed:1998rs,ArkaniHamed:1998nn,Antoniadis:1998ig,Randall:1999ee,Davoudiasl:1999tf,Huber:2000ie}. In particular, the universal extra dimensional (UED) model has been a subject of intense investigation recently because it can easily provide a dark matter candidate \cite{Servant:2002aq,Servant:2002hb,Bertone:2002ms}. This model, originally proposed in Ref.~\cite{Appelquist:2000nn}, is so named because all particles are allowed to propagate in the bulk and have universal access to all compact dimensions \cite{Appelquist:2000nn}. This is in contrast to both \mbox{Arkani-Hamed--Dimopoulos--Dvali} (ADD) type models where all standard model fields are restricted to a brane \cite{ArkaniHamed:1998rs} or models in which only some of the standard model fields can access the compact dimensions \cite{Muck:2001yv,Muck:2003kx}. A key issue when constructing a particle theory of dark matter is its stability on cosmological timescales. In the case of the UED models a vestigial discrete translation invariance,  known as Kaluza-Klein (KK) parity, ensures stability of the lightest KK particle (LKP) to all orders in perturbation theory. 

In this paper we discuss a theory containing many scalar dark matter states that originate from a single complex scalar in the factorized spacetime, $M_4 \times S^2$. This spherical compactification, with associated spherical-harmonic eigenfunctions labeled by quantum numbers $\ell$ and $m$, naturally organizes the associated four-dimensional KK states, $\tchi_\ell^m$, into towers of definite $m$, with  $\tchi_{|m|}^m$ being the lightest state in each tower.  These lightest KK particles are nominally stable due to the residual rotational symmetry about the polar axis of the extra dimensional geometry and an imposed global $U(1)$ symmetry.\footnote{Stable on cosmologically-interesting timescales.  While gravitational decay may occur we imagine here an effective theory with stabilized extra-dimensions and very massive shape moduli.}    This set of stable KK particles (SKPs) then comprises the set of dark matter states in the theory. 

Since this type of compactification does not allow for a chiral zero-mode fermion in the four-dimensional (4D) effective spectrum \cite{Camporesi:1995fb, Abrikosov:2001nj}, we restrict ourselves to a non-universal extra dimensional scenario.  For simplicity, we also restrict the entire standard model field content, not just the fermions, to the brane on $S^2$. As the position of the standard model brane on $S^2$ is completely arbitrary we can choose our coordinates such that it resides at the north pole.\footnote{We imagine here that the ultraviolet physics fixing the brane at the pole renders branon fluctuations very massive.}  While the location of the brane breaks some of the spherical rotational symmetry, rotations about the polar axis are preserved. Although we discuss here the simplest spherical case, we expect similar results will hold for any manifold which preserves a continuous isometry of the extra dimensional ground state metric.

In Section~\ref{sec:formalism} we describe the formalism of a complex scalar field on the $M_4 \times S^2$ spacetime while Section~\ref{sec:instability} discusses the instability of the excited KK particles and the stability of the SKPs within our effective framework. In Section~\ref{sec:abundance} we briefly describe the relic abundances of the corresponding SKPs and in section~\ref{sec:scat} we investigate the spin-independent nucleon scattering related to relevant direct-detection experiments. We make some final statements and discuss our conclusions in Section~\ref{sec:conc}.

\section{Formalism}
\label{sec:formalism}

We imagine a model where it is reasonable to approximate the extra dimensional geometry as $S^2$ below some ultraviolet scale where the extra-dimensional geometry is stabilized.  While in this work we assume an effective framework where the extra dimensional geometry is stabilized, there has been considerable interest in possible mechanisms for stabilization in the literature (See, e.g., Refs.~\cite{Sundrum:1998ns,Carroll:2003db,Carroll:2001ih}). 

We start by introducing the propagating complex scalar field, $\chi (x, \theta, \phi)$, onto the six dimensional (6D) spacetime, $M_4 \times S^2$, while the standard model fields are confined to a 3+1 brane at the north pole of the extra dimensions.  The entire 6D Lagrangian is then written as 
\begin{align}
\mathcal{L} = \mathcal{L}_{Bulk} + \delta \left( \cos \theta - 1 \right) \mathcal{L}_{Brane}
\end{align}
where the Dirac delta function serves to localize the field content of $\mathcal{L}_{Brane}$ to the north pole. The scalar field is assumed to be a singlet under the standard model gauge group but does have a global $U(1)$ symmetry associated with it both on and off the brane. The bulk Lagrangian is 
\begin{equation}
\mathcal{L}_{Bulk} = G^{AB}  \p_A \chi^* \p_B \chi - m_B^2 \left| \chi \right|^2 - \frac{g_1}{2×} \left( \left| \chi \right|^2 \right)^2 
\end{equation}
where $A,B=(0,1,2,3,4,5)$, $m_B$ is the 6D bulk mass, and the bulk coupling $g_1$ has mass dimension -2 in 6D. The metric, $G_{AB}$, is the 6D ground state metric for the $M_4 \times S^2$ spacetime whose line element is
\begin{equation}
ds^2 = \eta_{\mu \nu} dx^\mu dx^\nu - R^2 \left( d \theta^2 + \sin^2 \theta d \phi^2 \right)
\end{equation}
where $\eta_{\mu \nu}$ is the 4D Minkowski metric with signature $(1,-1,-1,-1)$ and $R$ is the (constant) radius of the extra dimensions. If we demand that the effective theory be renormalizable then, due to its gauge representation, the scalar is only allowed to interact with the standard model via a quadratic coupling to the Higgs on the brane \cite{Burgess:2000yq}. Since this interaction involves two spin zero bosons it necessarily modifies the Higgs potential.  However, by demanding the correct sign for the interaction term, the standard model vacuum structure is left unchanged. This leads to the brane Lagrangian
\begin{equation}
\mathcal{L}_{Brane} = \left( \mathcal{L}_{SM} - g_2 |\Phi|^2 |\chi|^2   \right) \delta \left( \cos \theta - 1 \right)
\end{equation}
where $\mathcal{L}_{SM}$ is the standard model Lagrangian and $g_2$ is the 6D brane coupling with mass dimension -2. Solving the free field equations in the bulk yields the harmonic decomposition of the scalar
\begin{equation}
\chi ( x, \theta, \phi ) = \frac{1}{R×} \sum_{\ell = 0}^\infty \sum_{m = - \ell}^{\ell} \tchi_\ell^m (x) Y_\ell^m (\theta, \phi)
\end{equation}
where the $\tchi_\ell^m$ fields are the 4D KK states. The factor of $R^{-1}$ is the necessary normalization to ensure that the KK states have the proper mass dimension for a scalar field in the 4D effective theory. As the sphere has no boundary, the boundary conditions are all trivially satisfied. Integrating over $S^2$ and using the orthogonality of the spherical harmonics we determine the total 4D effective bulk Lagrangian to be
\begin{align}
&\mathcal{L}_{Bulk}^{4D} = - \tchi_\ell^{* m} \left( \p_\mu \p^\mu + m_B^2 + \frac{1}{R^2×} \ell (\ell+1) \right) \tchi_\ell^m \notag \\ 
&- \frac{\overline{g}_1}{4 ×}  (-1)^{\overline{m}} B_{\;  \; \; \; \ell_1 \; \; \; \ell_2 \; \; \; \; \; \ell}^{-m_1 \; m_2 \; -m} B_{\;  \; \; \; \ell_3 \; \; \; \ell_4 \; \; \ell}^{-m_3 \; m_4 \; m} \tchi_{\ell_1}^{* m_1} \tchi_{\ell_2}^{m_2} \tchi_{\ell_3}^{* m_3} \tchi_{\ell_4}^{m_4} 
\end{align}
where $\overline{m} = m_1 + m_3 + m$ and $\overline{g}_1 = \displaystyle \frac{g_1}{4 \pi R^2×}$ is the (dimensionless) effective 4D bulk coupling for the scalar 4 point self interaction in the bulk. The free term implies the existence of a large mass degeneracy for the KK states. For each value of $\ell$, there exists $2 \ell + 1$ distinct KK states with the same mass, each denoted by the magnetic quantum number $m$ which is bounded as $-\ell \leq m \leq \ell$.  For each $m$ there exists a whole tower of KK states with $\ell \geq |m|$, and a lowest mass state with $\ell = |m|$. The symbol $ B_{\;  \ell_1 \; \; \ell_2 \; \; \ell_3}^{m_1 m_2 m_3}$ is defined as the weighted product of two Wigner 3-$j$ symbols
\begin{align}
B_{\;  \ell_1 \; \; \ell_2 \; \; \ell_3}^{m_1 m_2 m_3} &= \sqrt{(2 \ell_1 + 1) (2 \ell_2 + 1) (2 \ell_3 + 1)} \notag \\
& \times \left(
\begin{array}{ccc}
 \ell_1 & \ell_2 & \ell_3 \\
0 & 0 & 0
\end{array}
\right)
\left(
\begin{array}{ccc}
 \ell_1 & \ell_2 & \ell_3 \\
m_1 & m_2 & m_3 
\end{array}
\right)
\end{align}

The symmetries of the 3-$j$ symbols enforce conservation of angular momentum in the extra dimensions and impose certain selection rules on the bulk interactions. These symbols play a large role in understanding the instability of excited KK states. Note that, for brevity, we have adopted a summation convention such that for each pair of indices of the type $(m_i , \ell_i )$ appearing in the 4D effective theory we assume the presence of a corresponding summation operator $\sum_{\ell_i=0}^\infty \sum_{m_i=-\ell_i}^{\ell_i}$ unless stated otherwise. 

\begin{figure}[t]
\subfigure[The propagator for the KK states.]{\label{fig:FeynmanRules-a} \includegraphics[scale=0.45]{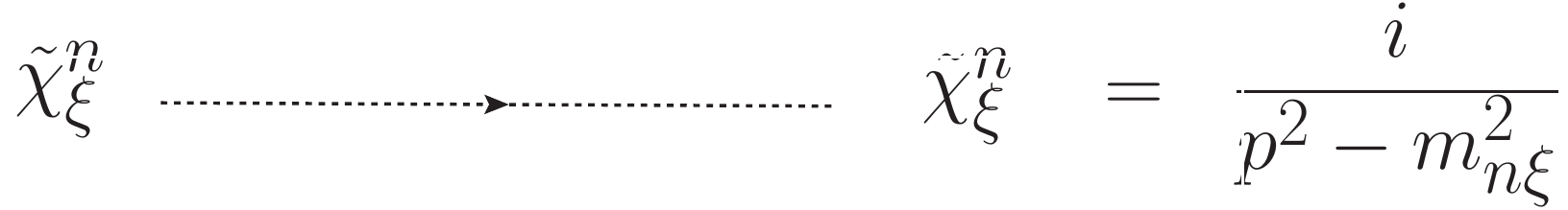}}
\subfigure[The three point interaction for non-magnetic KK states and the Higgs on the brane.]{\label{fig:FeynmanRules-b} \includegraphics[scale=0.385]{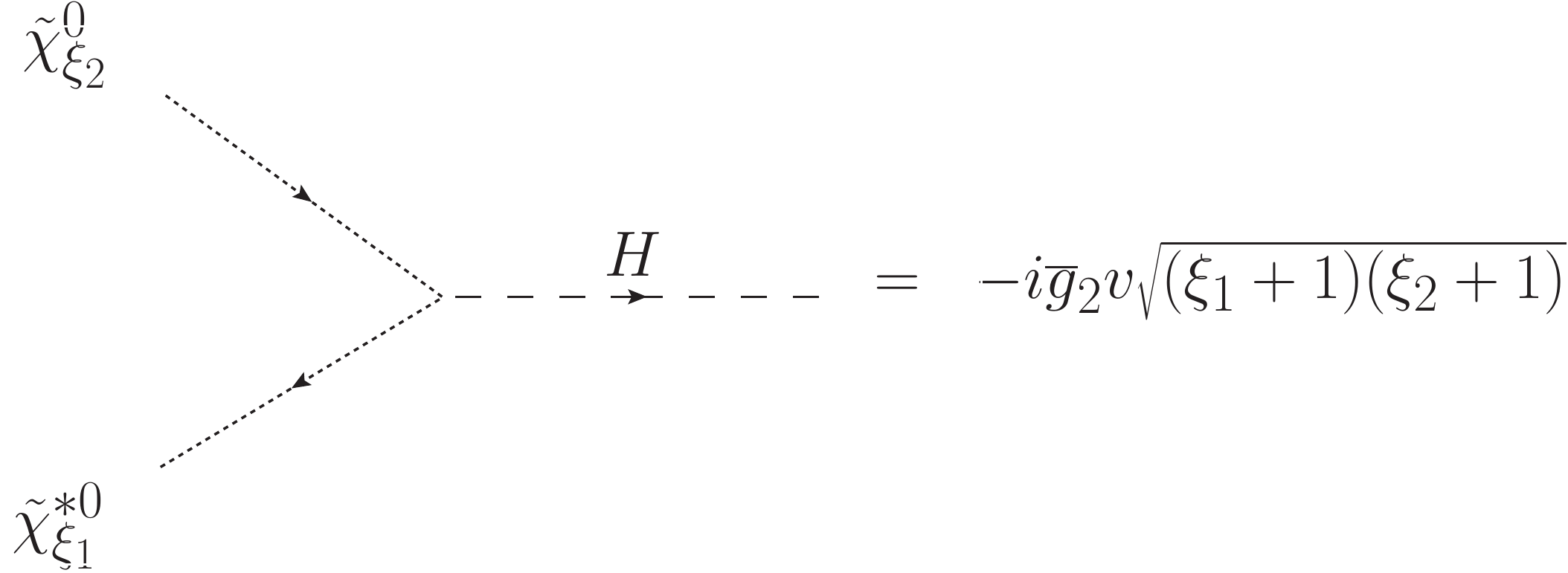}}
\subfigure[The four point interaction for non-magnetic KK states and Higgs fields on the brane.]{\label{fig:FeynmanRules-c} \includegraphics[scale=0.425]{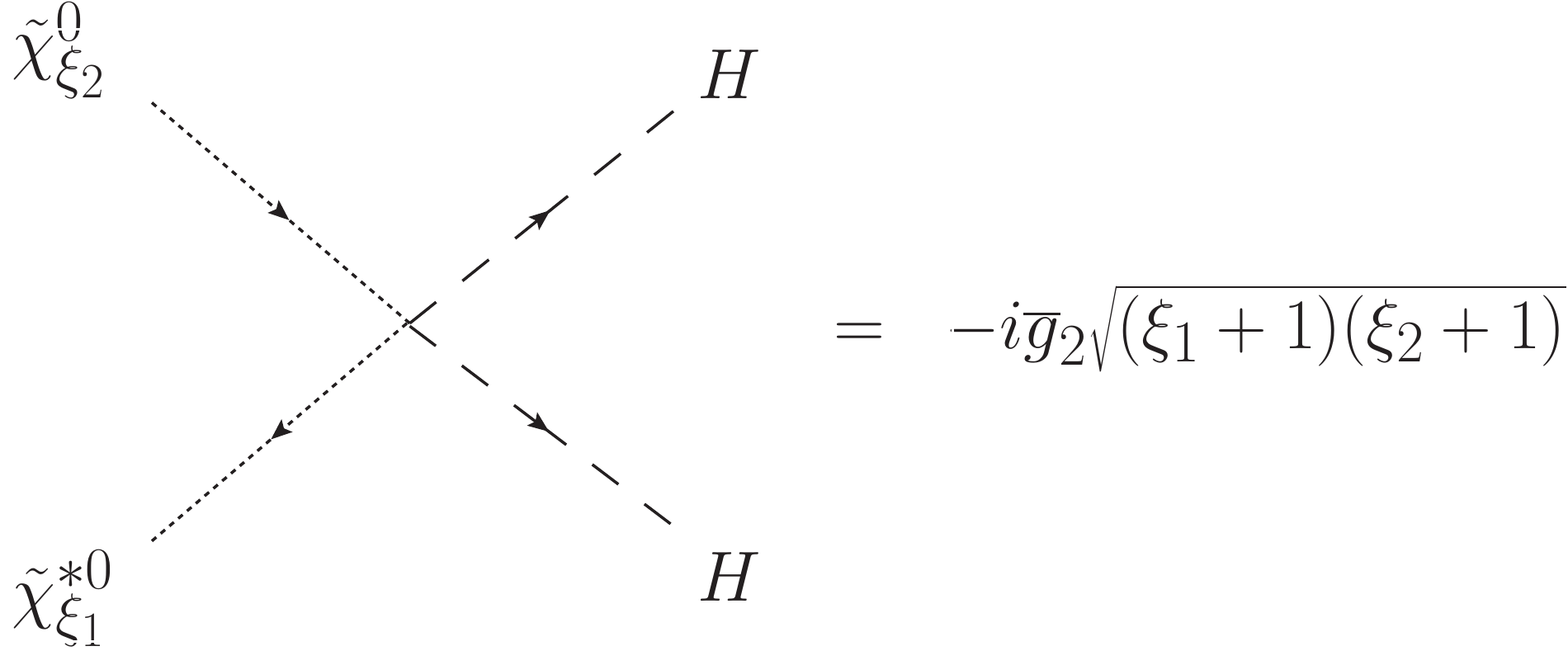}}
\subfigure[The four point interaction for magnetic and non-magnetic KK states in the bulk. Note that $\overline{n} = n_1 + n_3 + m$.]{\label{fig:FeynmanRules-d} \includegraphics[scale=0.45]{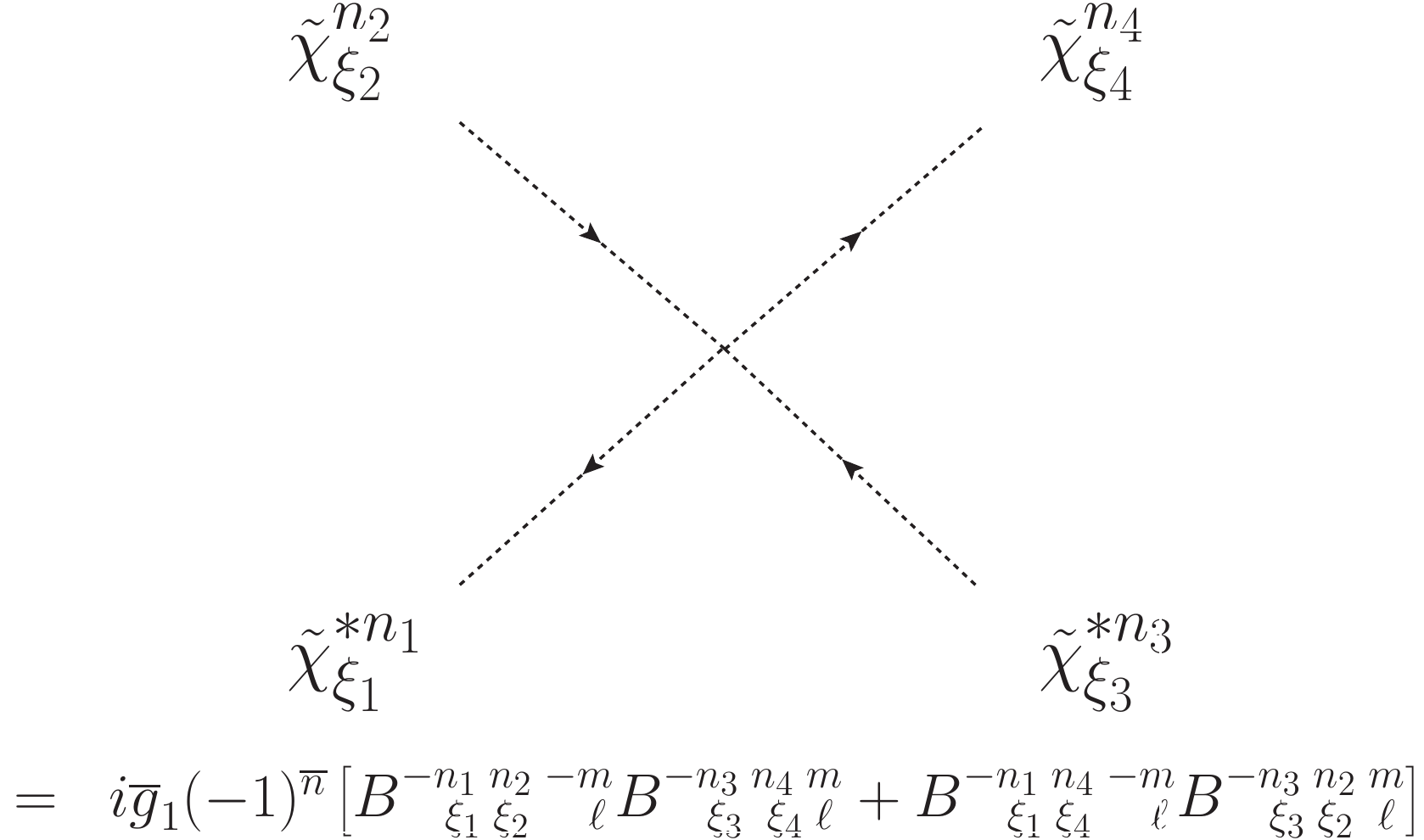}}
\caption{The feynman rules for the 4D effective theory.}
\label{fig:FeynmanRules}
\end{figure}

The effect of integrating over $S^2$ on the brane is to rescale the fields and couplings such that 

\begin{align}
\mathcal{L}_{Brane}^{4D} &= \mathcal{L}_{SM}^{4D} + \frac{1}{2×} \p_\mu H \p^\mu H  - \frac{\mu^2}{2×} H^2 - \frac{\mu v}{2×} H^3 \notag \\ 
& - \frac{\lambda}{8×} H^4 - \frac{\overline{g}_2}{2×} \left( H^2 + 2 v H  + v^2  \right) \overline{\tchi}_{\ell_1}^{* 0} \overline{\tchi}_{\ell_2}^{0}
\end{align}
where $v = \ds \sqrt{\frac{\mu}{2 \lambda×}}$ is the effective 4D Higgs vacuum expectation value (vev), $\ds \overline{g}_2 = \frac{g_2}{4 \pi R^2}$ is the effective 4D dimensionless brane coupling, and the scalar fields on the brane have been rescaled as $\overline{\tchi}_{\ell_i}^{0} = \sqrt{(2 \ell_i + 1)} \tchi_{\ell_i}^{0}$. Note that the pole localized interaction projects out the non-magnetic KK states, ($m_i=0$), on the brane while also breaking the spherical symmetry so that the total angular momentum quantum number, $\ell_i$, is no longer conserved. This non-conservation of total angular momentum on the brane generates an infinitely large mass mixing matrix for the non-magnetic modes. Since the brane mass, which is determined by diagonalizing the mixing matrix, is just a small perturbation of the bulk mass relation we can write the full 4D effective Lagrangian as
\begin{align}
\mathcal{L}^{4D} &=  - \tchi_\ell^{* m} \left( \p_\mu \p^\mu + m_{m \ell}^2 \right) \tchi_\ell^m + \mathcal{L}_{SM}^{4D} + \frac{1}{2×} \p_\mu H \p^\mu H \notag \\
& - \frac{\overline{g}_1}{4 ×}  (-1)^{\overline{m}} B_{\;  \; \; \; \ell_1 \; \; \; \ell_2 \; \; \; \; \; \ell}^{-m_1 \; m_2 \; -m} B_{\;  \; \; \; \ell_3 \; \; \; \ell_4 \; \; \ell}^{-m_3 \; m_4 \; m} \tchi_{\ell_1}^{* m_1} \tchi_{\ell_2}^{m_2} \tchi_{\ell_3}^{* m_3} \tchi_{\ell_4}^{m_4} \notag \\
& \; - \frac{\mu^2}{2×} H^2 - \frac{\mu v}{2×} H^3 - \frac{\lambda}{8×} H^4 - \frac{\overline{g}_2}{2×} \left( H^2 + 2 v H \right) \overline{\tchi}_{\ell_1}^{* 0} \overline{\tchi}_{\ell_2}^{0} 
\end{align}
where $m_{m \ell}^2 = m_B^2 + \ds \frac{1}{R^2×} \ell ( \ell + 1) + \delta_0^m \frac{\overline{g_2} v^2}{2×} ( 2 \ell + 1)$. The non-magnetic states therefore receive a small mass correction due to the electroweak symmetry breaking (EWSB) on the brane and the corrections to the eigenstates have been neglected since they are suppressed by a factor proportional to $\overline{g}_2 v^2 R^2$. Note that the correction from EWSB on the brane reduces the mass degeneracy by one so that, for each value of $\ell$, there is now only $2 \ell$ degenerate KK states. The Feynman rules for the 4D effective theory are shown in Figures~\ref{fig:FeynmanRules-a} through~\ref{fig:FeynmanRules-d}. \\

The use of a different set of labels for the quantum numbers, $(n_i,\xi_i)$ instead of $(m_i,\ell_i)$, indicates that there is no more implied summation associated with these particular values. The only implicit summations that appear in the feynman rules are over the $(m,\ell)$ quantum numbers associated with the effective four point self interaction in Figure~\ref{fig:FeynmanRules-d}. 

\section{Instability of Excited KK States}
\label{sec:instability}

In the present work, we are able to obtain stable states directly from the combination of the continuous rotational invariance about the polar axis and the imposed $U(1)$ symmetry. This method of stabilization leads naturally to a multi-state theory of dark matter. \\

There are two possible decay modes available to the KK states in the non-magnetic tower at the tree level given by $\tchi_{\xi_1}^0 \rightarrow \tchi_{\xi_2}^0 H$ and $\tchi_{\xi_1}^0 \rightarrow \tchi_{\xi_2}^0 H H$. The amplitudes for both of these modes are constant and imply that the total decay rate depends directly on the level of excitation of the KK states, i.e; the higher the excitation of the KK states the faster the rate becomes. There are no symmetry constraints on the amplitudes, as with the self interaction amplitude in the bulk, so the decays proceed unhindered as long as the relevant kinematics are satisfied. Since the imposed $U(1)$ symmetry prevents a decay mode for the lightest non-magnetic KK state, $\tchi_0^0$, into standard model particles, it constitutes our first SKP and lightest dark matter state. 

\begin{figure}[ht!]
\subfigure[]{\label{fig:1LoopDecays-a} \includegraphics[scale=0.45]{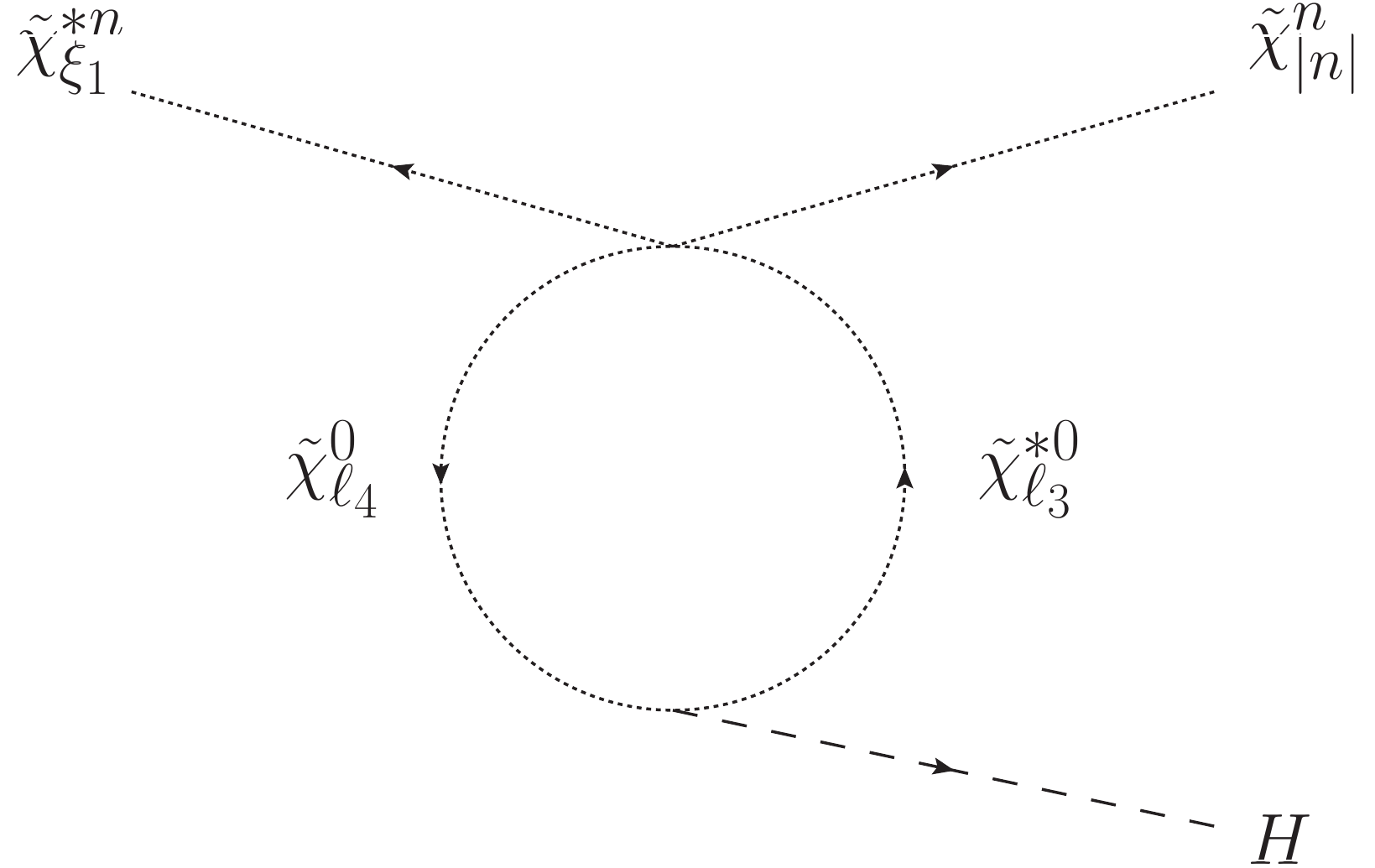}}
\subfigure[]{\label{fig:1LoopDecays-b} \includegraphics[scale=0.45]{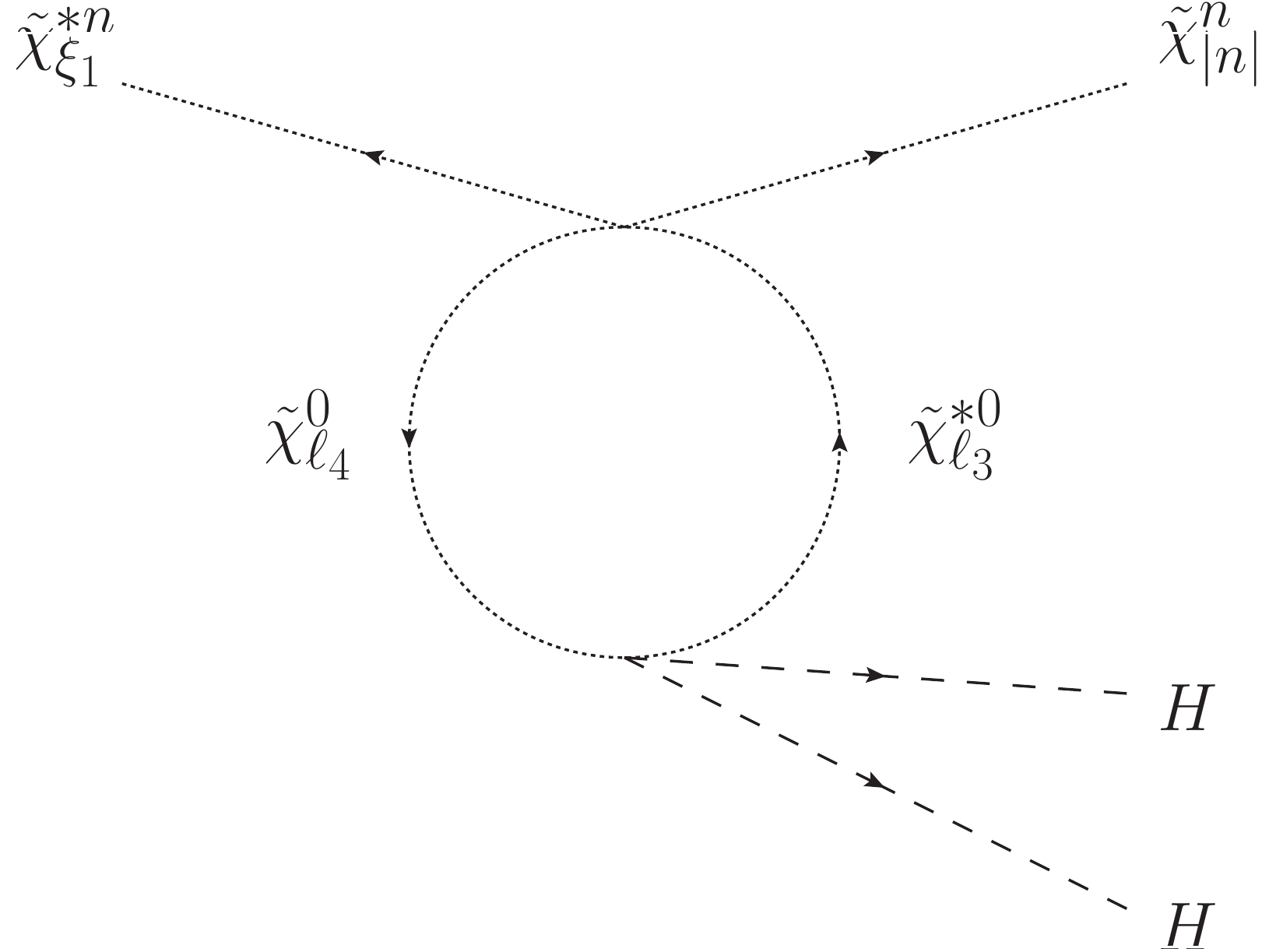}}
\caption{The one-loop decay channels which allow for the decay of the excited KK states in the bulk.}
\label{fig:1LoopDecays}
\end{figure}

At first glance, there are also many three-body decay modes available at the tree level for the magnetic states ($m_i \neq 0$) which are all of the general form $\tchi_{\xi_1}^{n_1} \rightarrow \tchi_{\xi_2}^{n_2} \tchi_{\xi_3}^{n_3} \tchi_{\xi_4}^{n_4}$. These decay modes are highly constrained by the selection rules imposed by the symmetries of the 3-$j$ symbols. In particular, the triangle inequalities place an upper bound on the $\xi_1$ quantum number such that $\xi_1 \leq \xi_2 + \xi_3 + \xi_4$. This imposes an upper limit on the mass of the decaying particle which prevents it from providing enough phase space to satisfy the necessary kinematics. Due to this, all of the decay modes occurring via the four point self interaction are in fact forbidden at the tree level. However, by considering the 1-loop corrections we can allow the excited magnetic states to decay via the two processes $\tchi_{\xi}^{n} \rightarrow \tchi_{|n|}^{n} H$ and $\tchi_{\xi}^{n} \rightarrow \tchi_{|n|}^{n} HH$ as shown in Figures~\ref{fig:1LoopDecays-a} and~\ref{fig:1LoopDecays-b} . The labels for the quantum numbers of the virtual KK states, ($0,\ell_i$), were chosen so as to imply the summation over all possible states which can participate in the loop. 

The total decay rate for a given excited KK state is then 

\begin{align}
\Gamma &= \frac{\overline{g}_1^2 \overline{g}_2^2}{4096 \pi^5 m_1} \left( M_1^2 + M_2^2 \right) \notag \\
& \times \left[ \sum_{\ell_3 = 0}^{\xi_{max}} \sum_{\ell_4 = 0}^{\xi_{max}} A_{\xi_1 \ell_3 \ell_4}^n \ln \left( \frac{\xi_{max} (\xi_{max} + 1)}{R^2 m_{0 \ell_3}^2} \right) \right]^2
\end{align}

where 
\setcounter{equation}{9}
\begin{subequations}
\begin{equation}
M_1^2 = \frac{v^2 \sqrt{(m_1^2 + m_2^2 - m_H^2)^2 4 m_1^2 m_2^2}}{m_1^2}
\end{equation}
\begin{equation}
M_2^2 = \frac{(m_1 - m_2 - 2 m_H)^2}{16 \pi^2}
\end{equation}
and 
\begin{align}
&A_{\xi_1 \ell_3 \ell_4}^n = \sqrt{(2 \ell_3 + 1)(2 \ell_4 + 1)} (-1)^n \notag \\
& \times \left[ B_{\; \; \; \xi_1 \; |n| \; \ell}^{-n \; \; \; n \; \; 0} \; B_{\; \ell_3 \;  \ell_4  \; \; \ell}^{\; 0 \; \; \; 0 \; \; 0} + (-1)^{n} B_{\; \; \; \xi_1 \; \; \ell_4 \; \ell}^{-n \; \; \; 0 \; \; n} \; B_{\; \ell_3 \;  |n|  \; \; \; \;\ell}^{\; 0 \; \; \; n \; \; -n}\right]
\end{align}
\end{subequations}

We have assumed that the theory holds up to some mass scale which can be written in the general form $\Lambda^2 = \xi_{max}(\xi_{max}+1)/R^2$. The $\tchi_{|n|}^{|n|}$ state is the SKP belonging to the same KK tower as the original excited KK state $\tchi_{\xi_1}^n$. The fact that the SKP in the end state must belong to the same KK tower as the original decaying excited KK state is insured by the particular symmetry of the 3-$j$ symbols that corresponds to the rotational invariance about the polar axis. The SKP of each magnetic KK tower is then preserved by the same continuous rotational symmetry. We conclude that any excited KK state will decay, either through tree level or 1-loop processes, down to the SKP of its respective tower which is then stable. This theory then naturally behaves as a multi-state theory of dark matter with the SKPs as the set of dark-matter candidates.

\section{Dark Matter Relic Abundance}
\label{sec:abundance}

In theories where dark-matter is produced thermally the appropriate Boltzmann equation for number-changing processes must be solved to show that the parameter space can accomodate a cosmological relic abundance of $\Omega_d h^2 \simeq 0.11$ to match observations \cite{Dunkley:2008ie}. The model discussed here has a four dimensional parameter space, $(\overline{g}_1,\overline{g}_2,m_B,R)$, corresponding to the effective bulk and brane coupling strengths, the bulk mass, and the length scale of the extra dimensions. The relic densities of the SKPs are determined by solving the set of coupled Boltzmann equations given by
\begin{align}
\frac{d \beta_{n_1 |n_1|}}{ dx ×} &= \sqrt{ \frac{\pi}{45}} \frac{g^s_*(T)}{\sqrt{g_*(T)}}\frac{M_{Pl} m_0}{x^2×} \Bigg[ \beta^{eq}_{ n_1 |n_1|} \beta^{eq}_{m_2 \ell_2} \; \Delta   \tcs^{n_1 m_2 m_3 m_4}_{\ell_2 \ell_3 \ell_4} \notag \\ \notag \\
& \times \left( \frac{\beta_{m_3 \ell_3} \beta_{m_4 \ell_4}}{\beta_{m_3 \ell_3}^{eq} \beta_{m_4 \ell_4}^{eq}×} - \frac{\beta_{n_1 |n_1|} \beta_{m_2 \ell_2}}{\beta_{n_1 |n_1|}^{eq} \beta_{m_2 \ell_2}^{eq}×} \right) \notag
\end{align}
\begin{equation}
+ \delta_0^{n_1} \Delta' \sum_{SM} \tcs_{SM}^{\ell_2} \bigg( \beta_{n_1 |n_1|}^{eq} \beta_{m_2 \ell_2}^{eq} - \beta_{n_1 |n_1|} \beta_{m_2 \ell_2} \bigg) \Bigg]
\end{equation}
where $M_{Pl}$ is the Planck mass, $g_*(T)$ and $g^s_*(T)$ are the effective number of density and entropy degrees of freedom in the thermal bath at temperature $T$, $m_0$ is the mass of the lightest SKP $\tchi_0^0$, and $x=m_0/T$ is an evolution variable.  The $\beta_{m_i \ell_i}$ variables are the number density to entropy ratios.
The $n_1$ index is a vector index labelling the different Boltzmann equations and we have again made use of the implicit summation convention to sum over the 6 quantum numbers associated with $\beta_{m_2 \ell_2}$, $\beta_{m_3 \ell_3}$, and $\beta_{m_4 \ell_4}$. These summations insure that we account for all possible interactions that the $n_1^{th}$ SKP can participate in. The coupling terms occur each time the summations produce a term which implies a 4 point interaction in which multiple SKPs are participatory. The thermalized cross sections are also dependent on all 7 quantum numbers via the amplitudes and by defining $\Delta$ and $\Delta'$ as 
\begin{align}
\Delta =
\left\{
\begin{array}{lr}
1 & \textrm{if } |m_3| + \ell_3 + |m_4| + \ell_4 - |m_2| - \ell_2 - 2 |n_1| \neq 0 \\
0 & \textrm{if } |m_3| + \ell_3 + |m_4| + \ell_4 - |m_2| - \ell_2 - 2 |n_1| = 0
\end{array}
\right.
\end{align}
and
\begin{align}
\Delta' = \delta_0^{m_2} \delta_0^{m_3} \delta_0^{\ell_3} \delta_0^{m_4} \delta_0^{\ell_4}
\end{align}
we filter out any non number-changing interactions and prevent any over counting. The magnetic SKPs interact solely through the four point functions originating in the bulk and are governed by the first term only. The non-magnetic SKP interacts directly with the Higgs field only but can undergo scattering processes with other standard model particles via Higgs exchanges. The sum $\sum_{SM}$ is meant to represent a summation over all possible allowed standard model processes, i.e; interactions of the following types $\tchi_0^{* 0} \tchi_\ell^0 \rightarrow \overline{f} f$, $\tchi_0^{* 0} \tchi_\ell^0 \rightarrow Z Z$, $\tchi_0^{* 0} \tchi_\ell^0 \rightarrow W^+ W^-$, and $\tchi_0^{* 0} \tchi_\ell^0 \rightarrow H H$. The simultaneous solution of these coupled differential equations leads directly to the desired set of SKP relic densities. The details and parameter dependence of these solutions will appear elsewhere \cite{Winslow&Sigurdson}.  Here, we simply mention that the resulting solutions confirm that viable regions of the parameter space exist.

\begin{figure}[h]
\includegraphics[width=3.35in, angle=0]{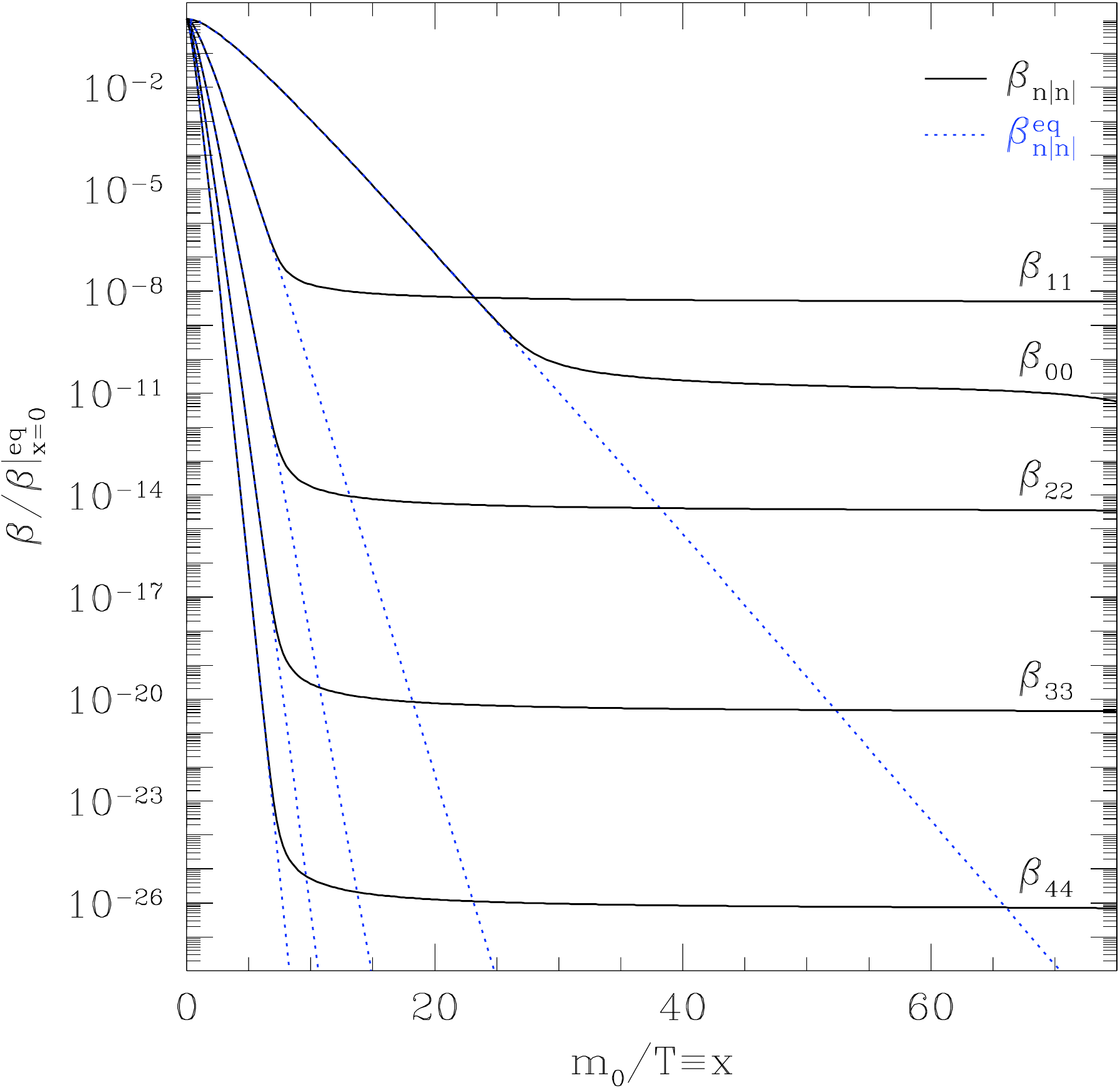}
\caption{The solutions for the first five states of the coupled Boltzmann equations that determine the relic abundance, along with their corresponding equilibrium values. In this case the parameters were chosen so that the $n_1 \neq 0$ states dominate the relic density.}
\label{fig:5Solutions}
\end{figure}

The contribution to the total dark-matter density from a given SKP is given by 

\begin{equation}
\Omega_{d,n_1} = \frac{g^s_0 m_{n_1 |n_1|} T_0^3}{\rho_{\rm c0}} \beta_{n_1 |n_1|}
\end{equation}
where $T_0$ is the present temperature of the Universe, $g^s_0 \simeq 3.91$, and $\rho_{\rm c0}$ is the critical density of the Universe today. The total contribution from all the SKPs is then just $\Omega_{d} = \sum_{n_1} \Omega_{d,n_1}$. In Figure~\ref{fig:5Solutions} we show the solution to the Boltzmann equations for the first five SKPs. 

The solid black lines represent the actual solutions of the Boltzmann equation while the dashed blue lines represent their corresponding equilibrium abundances. The mass parameters and effective couplings here were chosen so as to maximize the contribution from the magnetic SKPs while minimizing the contribution from the non-magnetic SKP.  We see that, in this example, only the $|n_1|=1$ SKP state contributes significantly while the $|n_1|>1$ SKP states remain significantly suppressed despite their enhancement due to the specific choice of parameters.

As will be discussed in Ref.~\cite{Winslow&Sigurdson}, since it annihilates into standard model states, we can increase the contribution of the lightest SKP with $\ell=|m|=0$ independently from the rest by decreasing the brane coupling.  The above example then implies that it is typically only the two lightest SKPs that yield significant contributions to the dark-matter density of the Universe. We therefore consider only the two lightest SKPs when investigating the consequences for direct detection experiments in the next section.

\section{Spin-Independent Nucleon Scattering}
\label{sec:scat}

Dark matter is distributed throughout the halo hosting the Milky Way and therefore might scatter off of target nuclei within detectors here on Earth as we move through the halo~ (e.g., Ref.~\cite{Jungman:1995df}).  A number of experiments, based on identifying nuclear recoil events corresponding to dark matter nucleon scattering, are currently in progress. 

\begin{figure}[t]
\subfigure[]{\label{fig:ScalarNucleonScattering-a} \includegraphics[scale=0.4]{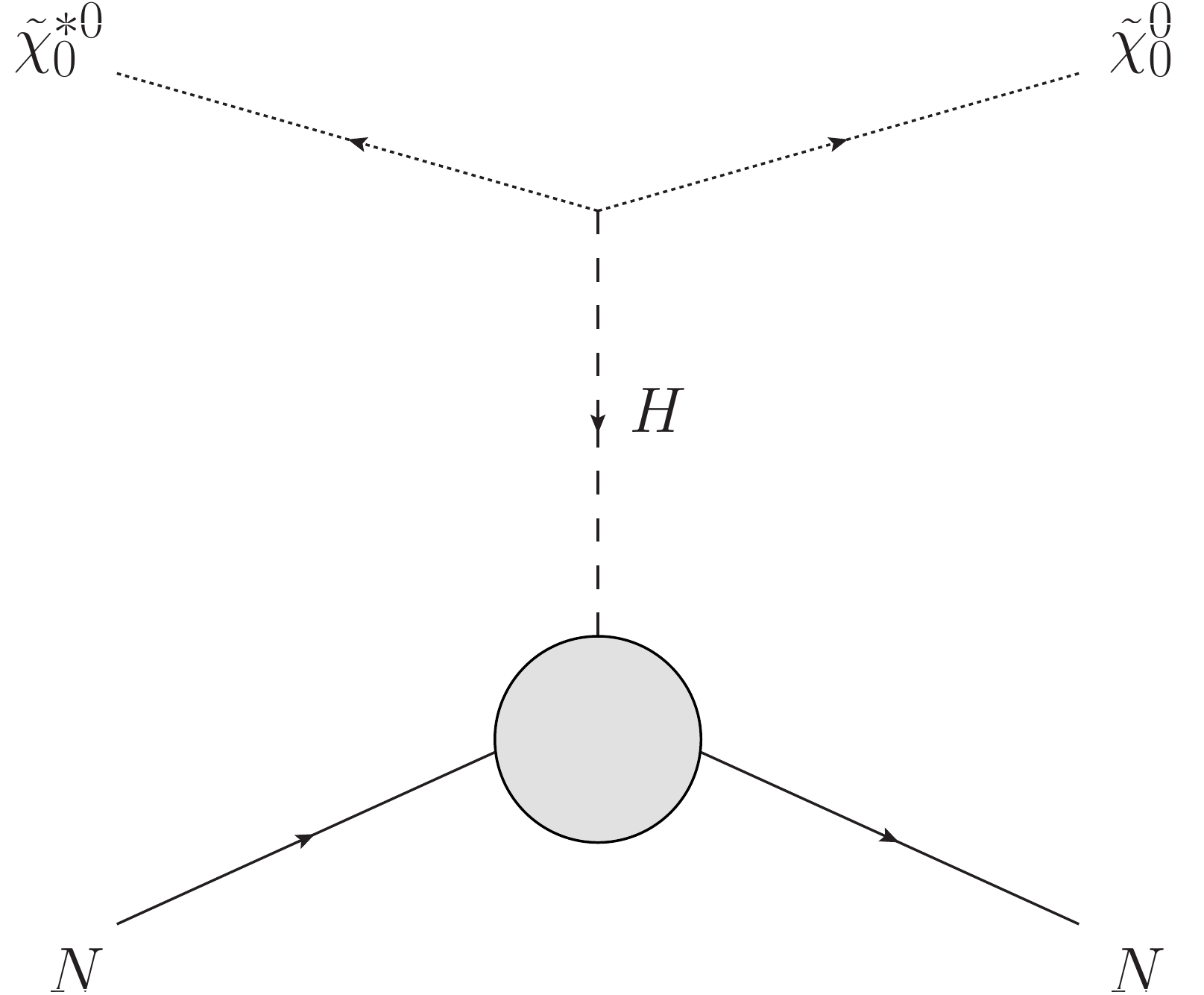}}
\subfigure[]{\label{fig:ScalarNucleonScattering-b} \includegraphics[scale=0.4]{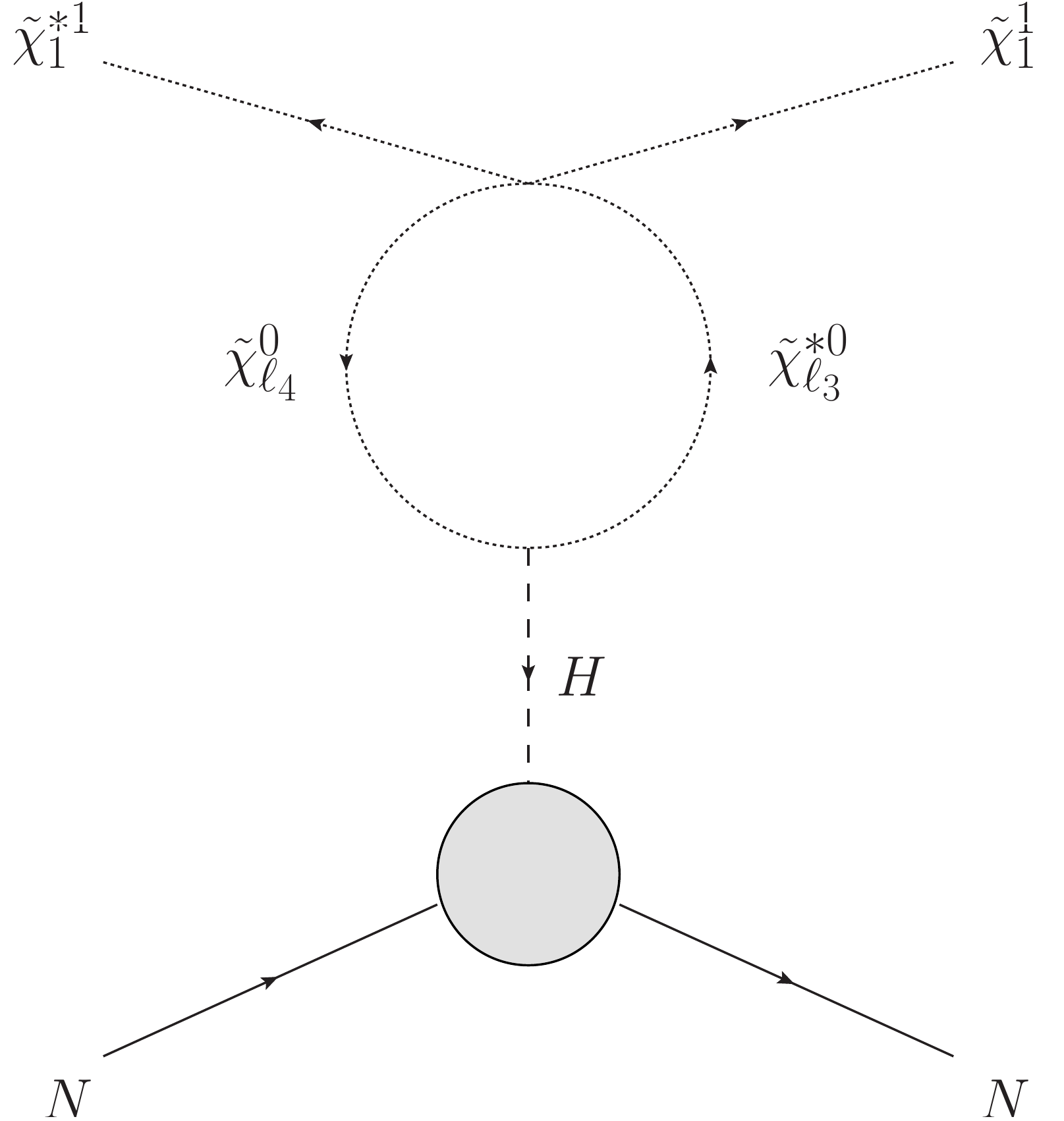}}
\caption{The effective diagrams contributing to spin-independent nucleon elastic scattering for the two lightest SKPs.}
\label{fig:ScalarNucleonScattering}
\end{figure}

We investigate here spin-independent elastic scattering of the two lightest SKPs with nucleons. The two relevant processes are shown in Figures~\ref{fig:ScalarNucleonScattering-a} and~\ref{fig:ScalarNucleonScattering-b}. The effective Higgs-nucleon Yukawa coupling includes both the direct coupling to the light quarks as well as an enhancement factor due to the introduction of heavy quark loops which contribute to the mass of the nucleon through the anomaly~\cite{Shifman:1978zn,Vainshtein:1980ea}. The relevant microscopic Lagrangian for the Higgs-nucleon coupling is given by

\begin{align}
\mathcal{L} = \sum_q \frac{m_q}{v} H \overline{q} q
\end{align}

The nucleonic matrix elements of the light-quark currents are obtained in chiral perturbation theory from measurements of the pion-nucleon sigma term \cite{Cheng:1988cz,Cheng:1988im,Gasser:1990ce}

\begin{align}
\langle N| m_q \overline{q} q | N \rangle = m_N f^{(N)}_{Tq}
\end{align}

while the nucleonic matrix elements of the heavy quarks are given by \cite{Jungman:1995df} 

\begin{align}
\langle N| m_Q \overline{Q} Q | N \rangle = \frac{2}{27×} m_N \left(1 - \sum_{q=u,d,s} f^{(N)}_{Tq} \right)
\end{align}

The above results imply the effective Higgs-nucleon coupling

\begin{equation}
\mathcal{V} = \frac{2 m_N}{9 v×} \left( 1 + \frac{7}{2×} \sum_{q=u,d,s} f^{(N)}_{T q} \right)
\end{equation}
where we employ the following estimates of the proton parameters \cite{Cheng:1988im,Gasser:1990ce} $f^{(p)}_{Tu}=0.019$, $f^{(p)}_{Td}=0.041$, $f^{(p)}_{Ts}=0.014$ and neutron parameters $f^{(n)}_{Tu}=0.023$, $f^{(n)}_{Td}=0.034$, $f^{(n)}_{Ts}=0.014$. \\

The total cross section for $\tchi_0^0$-$N$ scattering is given by
\begin{align}
\sigma_0 = \frac{\overline{g}_2^2 v^2 m_N^2 \mathcal{V}^2}{4 \pi m_H^4 m_0^2×} 
= 3.8 \times 10^{-39} \; \overline{g}_2^2 \; \bigg( \frac{\textrm{GeV}}{m_0} \bigg)^2 \textrm{cm}^2
\end{align}

where we have assumed a Higgs mass of 150 GeV. \\

\begin{figure}[h]
\includegraphics[width=3.35in, angle=0]{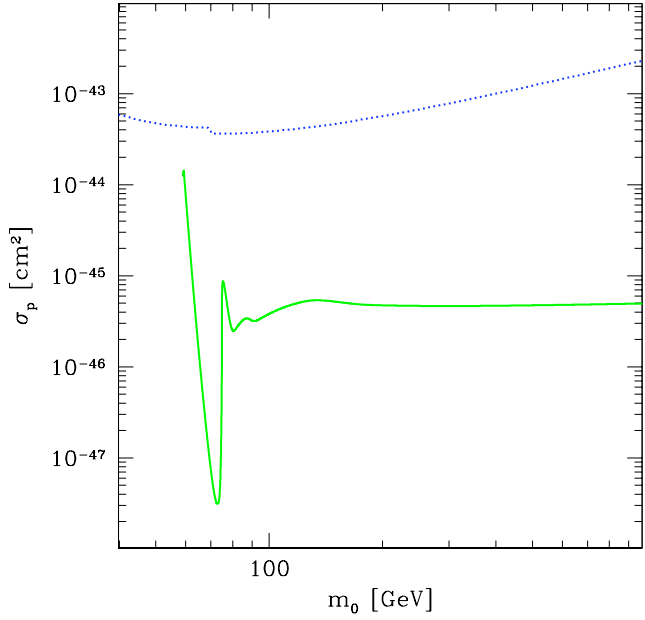}
\caption{The upper dashed curve (blue) is the current combined limit from the CDMSII collaboration.  The lower solid curve (green) shows the scattering cross section of the $\tchi_0^0$ state assuming it has a relic abundance near $\Omega_d h^2 \simeq 0.11$.
The $\tchi_0^0$ mass remains unconstrained by the current combined limit from the CDMSII collaboration.}
\label{fig:ComparisonOfSingleComponentTheory}
\end{figure}

In Figure~\ref{fig:ComparisonOfSingleComponentTheory} we compare the total cross section for $\tchi_0^0$-proton scattering with the recently released results from the CDMSII collaboration as reported in~\cite{Ahmed:2009zw}. The upper (blue) curve shows the combined limit of the full data set recorded at Soudan to date. The region above this line has been excluded. The lower curve is the total $\tchi_0^0$-proton cross section for parameters such that the non-magnetic SKP dominates the total dark matter relic abundance. 

In this limit the values of the brane coupling $\overline{g}_2$ and the bulk mass $m_B$ are restricted such that the relic abundance of the non-magnetic SKP satisfies the constraint $\Omega_d h^2 = \Omega_{d,0} h^2 \simeq 0.11 $. The behaviour of the $\tchi_0^0$-proton scattering cross section can be understood in view of this result. At late times the solution to the single state Boltzmann equation $\beta_{00}$ is inversely proportional to the total thermalized cross section $\tcs_{SM}$. Although a temperature dependent thermalized cross section is possible this leads only to slight numerical changes~\cite{Jungman:1995df} and we can therefore take it to be approximately constant. Since $\tcs_{SM}$ is proportional to the ratio $\ds \frac{\overline{g}_2^2}{m_0^2}$ the relic abundance constraint will associate an approximately constant value to this ratio and therefore to the $\tchi_0^0$-proton scattering cross section. 

Note that the decrease in the viable $\tchi_0^0$-proton cross section for low values of $m_0$ can be understood by virtue of the fact that the dominant processes contributing to $\tcs_{SM}$ are mediated by Higgs exchange so the relevant amplitudes contain a factor of a Higgs propagator. This implies that $\Omega_d \sim \ds {m_0^2 (s - m_H^2)^2}{/{\overline{g}}_2^2}$ or $\sigma_0 \sim (s - m_H^2)^2$. In the non-relativistic regime the transfer momentum $s$ is proportional to $4 m_0^2$ in the center of mass frame so we should expect that when $m_0 \sim m_H/2$ the virtual Higgs goes on-shell and we need to employ the Breit-Wigner resonance correction $(s - m_H^2)^2 \rightarrow (s - m_H^2)^2 + m_H^2 \Gamma_H^2$ where $\Gamma_H$ is the total Higgs width. Within the region where the Higgs is approximately on shell, $m_0 \sim 75$ GeV, we should then expect a decrease in the $\tchi_0^0$-proton scattering cross section.

The relic abundance constraint also prevents the brane coupling from taking on exceedingly low values as the bulk mass is lowered down, e.g., when $m_B=0$ the brane coupling is constrained such that the minimum mass value for the non-magnetic SKP is $m_0 \simeq 60$ GeV. This then explains the subsequent increase and eventual abrupt termination of the $\tchi_0^0$-proton cross section for low values of $m_0$ after the Higgs goes back off shell. The end result is that the combined limit from the recently released CDMSII results cannot meaningfully constrain the non-magnetic SKP mass. \\

The total cross section for $\tchi_1^1$-$N$ scattering is given by 

\begin{equation}
\sigma_1 = \frac{\overline{g}_1^2 \overline{g}_2^2 v^2 m_N^2 \mathcal{V}^2}{(4 \pi)^5 m_H^4 m_{11}^2} \left[ \sum_{\ell_3,\ell_4 = 0}^{\xi_{max}} A_{1 \ell_3 \ell_4}^1 \ln \left( \frac{\xi_{max} (\xi_{max} + 1)}{R^2 m_{0 \ell_3}^2} \right) \right]^2
\end{equation}

Due to the fact that $\tchi_1^1$-$N$ scattering must proceed via an extra $\tchi_\ell^0$ loop factor in the amplitude and the fact that the $\tchi_1^1$ mass has a contribution from the compactification radius, the cross section for $\tchi_1^1$-$N$ scattering is considerably lower than that for $\tchi_0^0$-$N$ scattering. Unfortunately, this implies that the existing direct detection limits cannot meaningfully constrain the $\tchi_1^1$ mass either. \\

For completeness we briefly discuss the possibility of inelastic scattering. In this case, the relevant scattering processes are $\tchi_0^0 N \rightarrow \tchi^0_{\xi} N$ and $\tchi_1^1 N \rightarrow \tchi^1_{\xi} N$. In each case, the final state particle is chosen to be the first excited KK state within the respective KK tower in order to minimize the mass splitting which we denote as $\delta=m_\xi - m_{DM}$. In its current minimal form the present theory only contains interactions for which the mass splittings are positive. The kinematics restricts the maximum recoil energy to be~\cite{Petriello:2008jj}

\begin{align}
E_{max} = \frac{\mu_1}{M_N×} \left[ \mu_1 u^2 - \delta + \sqrt{\mu_1^2 u^4 - 2 \mu_1 \delta u^2} \right]
\end{align}

where $M_N$ is the mass of the nucleus, $u$ is the incoming dark matter velocity, and $\mu_1$ is the reduced mass of the nucleus and the incoming dark matter $\mu_1=m_{DM} M_N/(m_{DM} + M_N)$. This restriction constrains the mass splittings such that $\delta < \mu_1 u^2/2$. Since typical bounds on the mass splittings from the various direct detection experiments tend to be $\sim \mathcal{O}(10)$ keV this implies that the inelastic scattering processes are only physically relevant in the large $R$ limit. In this limit the total cross sections for the two processes listed above are given respectively by 

\begin{align}
\sigma_0 = \frac{\overline{g}_2^2 v^2 m_N^2 \mathcal{V}^2 (2 \xi + 1)}{4 \pi m_H^4 m_{0}^2×} \sqrt{1 - \frac{\left( 2 \lambda \delta (m_N + E_{DM}) - \lambda^2 \delta^2 \right)}{|\vec{p}_{DM}|^2}}
\end{align}

\begin{align}
\sigma_1 &= \frac{\overline{g}_1^2 \overline{g}_2^2 v^2 m_N^2 \mathcal{V}^2}{(4 \pi)^5 m_H^4 m_{11}^2} \left[ \sum_{\ell_3,\ell_4 = 0}^{\xi_{max}}  A^1_{\xi \xi_3 \xi_4} \ln \left( \frac{\xi_{max}(\xi_{max}+1)}{R^2 m_3^2} \right) \right]^2 \notag \\ 
& \times \sqrt{1 - \frac{\left( 2 \lambda \delta (m_N + E_{DM}) - \lambda^2 \delta^2 \right)}{|\vec{p}_{DM}|^2}}
\end{align}

where $\vec{p}_{DM}$ and $E_{DM}$ are the incoming dark matter momentum and energy respectively and 

\begin{align}
\lambda = \frac{(m_\xi + m_{DM})}{2 m_N}
\end{align}
 
As a final remark, we attempt to answer the question: how might we know if there is more than one type of dark matter particle? Suppose we are able to measure the $\tchi_0^0$-$N$ cross section to obtain the value of the brane coupling as well as the $\tchi_0^0$ mass. We can then calculate the relic density for $\tchi_0^0$ and compare this value with the WMAP observation. If the value comes up short we can interpret this as a signal that there exists at least one other dark matter state whose scattering cross section with the corresponding nucleon is suppressed by various loop factors and a higher mass but it, nonetheless, makes a significant contribution to the total dark matter relic abundance. 

\section{Conclusions}
\label{sec:conc}
We have demonstrated a novel particle model which uses an ADD type braneworld scenario to produce a multi-state theory of dark matter. Compactification of the extra dimensions onto a sphere leads to the association of a single complex scalar bulk field with multiple KK towers in the 4D effective theory. The lightest KK state in each tower was then naturally stabilized by the combination of an imposed $U(1)$ symmetry and a subset of the continuous spherical symmetry  of the extra dimensions(though much of the formalism we discuss here would remain true even if these states were meta-stable on cosmological timescales). We have also shown that viable regions of the parameter space exist that can readily produce the observed dark-matter relic abundance~\cite{Winslow&Sigurdson}. This model  remains unconstrained after comparison of the spin-independent $\tchi_0^0$-proton cross section with recent direct detection data. The most important aspect of this paper is that the multi-state nature of the dark matter is a direct and natural consequence of the continuous compactification geometry. 

\acknowledgements We would like to thank C.P. Burgess, K. McDonald, M.V. Raamsdonk, and B. Underwood for useful discussions. The combined spin-independent limit from the CDMSII direct detection experiment was obtained from the http://dmtools.brown.edu website operated by R. Gaitskell and J. Filippini. This work was supported in part by a Natural Sciences and Engineering Research Council (NSERC) of Canada Discovery Grant. PW was also supported in part by NSERC.

\end{document}